\documentclass[12pt]{article}
\usepackage{amssymb}

\input{tcilatex}
\begin{document}

\bigskip \ 

\bigskip \ 

\begin{center}
\textbf{A LINK BETWEEN BLACK HOLES}

\bigskip \ 

\textbf{AND THE GOLDEN RATIO}

\bigskip \ 

\smallskip

J. A. Nieto\footnote{%
nieto@uas.uasnet.mx, janieto1@asu.edu}

\smallskip

\textit{Facultad de Ciencias F\'{\i}sico-Matem\'{a}ticas de la Universidad
Aut\'{o}noma}

\textit{de Sinaloa, 80010, Culiac\'{a}n Sinaloa, M\'{e}xico}

\smallskip

and

\smallskip

\textit{Mathematical, Computational \& Modeling Sciences Center, Arizona
State University, PO Box 871904, Tempe AZ 85287, USA}\bigskip \ 

\bigskip \ 

\textbf{\bigskip \ }

\textbf{Abstract}
\end{center}

We consider a variational formalism to describe black holes solution in
higher dimensions. Our procedure clarifies the arbitrariness of the radius
parameter and, in particular, the meaning of the event horizon of a black
hole. Moreover, our formalism enables us to find a surprising link between
black holes and the golden ratio.

\bigskip \ 

\bigskip \ 

\bigskip \ 

\bigskip \ 

\bigskip \ 

\bigskip \ 

\bigskip \ 


Pacs numbers: 04.60.-m, 04.65.+e, 11.15.-q, 11.30.Ly

June, 2011

\newpage

It is well known that the ansatz for a spherically symmetric static black
hole solution in a $d$-dimensional space-time $M^{d}$ can be written as (see
Refs. [1]-[5] and references therein)

\begin{equation}
\gamma _{\mu \nu }=\left( 
\begin{array}{ccc}
-e^{f(r)} & 0 & 0 \\ 
0 & e^{h(r)} & 0 \\ 
0 & 0 & \varphi ^{2}(r)\tilde{\gamma}_{ij}(\xi ^{k})%
\end{array}%
\right) ,  \tag{1}
\end{equation}%
where $\tilde{\gamma}_{ij}(\xi ^{k})$ determines a maximally spherically
symmetric space in ($d-2)$ dimensions. From this metric one finds that the
only non-vanishing Ricci tensor components are (see Ref. [6]-[8] and
references therein)

\begin{equation}
R_{11}=e^{f-h}(\frac{\ddot{f}}{2}+\frac{\dot{f}^{2}}{4}-\frac{\dot{f}\dot{h}%
}{4}+\frac{(d-2)}{2}\dot{f}\frac{\dot{\varphi}}{\varphi }),  \tag{2}
\end{equation}

\begin{equation}
R_{22}=-\frac{\ddot{f}}{2}-\frac{\dot{f}^{2}}{4}+\frac{\dot{f}\dot{h}}{4}+%
\frac{(d-2)}{2}\dot{h}\frac{\dot{\varphi}}{\varphi }-(d-2)\frac{\ddot{\varphi%
}}{\varphi },  \tag{3}
\end{equation}%
\begin{equation}
R_{ij}=e^{-h}\{ \frac{(\dot{h}-\dot{f})\varphi \dot{\varphi}}{2}-\varphi 
\ddot{\varphi}-(d-3)\dot{\varphi}^{2}\} \tilde{\gamma}_{ij}+k(d-3)\tilde{%
\gamma}_{ij}.  \tag{4}
\end{equation}%
Here, we used the notation $\dot{A}\equiv \frac{dA}{dr}$ for any function $%
A=A(r)$. Moreover, we have $k=\{-1,0,1\}$. Thus, the Ricci scalar $R=\gamma
^{\mu \nu }R_{\mu \nu }$ becomes

\begin{equation}
\begin{array}{c}
R=e^{-h}\{-\ddot{f}-\frac{\dot{f}^{2}}{2}+\frac{\dot{f}\dot{h}}{2}+(d-2)(%
\dot{h}-\dot{f})\frac{\dot{\varphi}}{\varphi }-2(d-2)\frac{\ddot{\varphi}}{%
\varphi } \\ 
\\ 
-(d-3)(d-2)\frac{\dot{\varphi}^{2}}{\varphi ^{2}}\}+k(d-3)(d-2)\varphi ^{-2}.%
\end{array}
\tag{5}
\end{equation}

On the other hand, we have

\begin{equation}
\sqrt{-\gamma }=e^{\frac{f+h}{2}\varphi ^{(d-2)}}\sqrt{\tilde{\gamma}}, 
\tag{6}
\end{equation}%
where $\gamma $ and $\tilde{\gamma}$ denote the determinant of $\gamma _{\mu
\nu }$ and $\tilde{\gamma}_{ij}$, respectively. Consequently, up to total
derivative, the higher dimensional Einstein-Hilbert action

\begin{equation}
S=\frac{1}{2}\int_{M^{d}}\sqrt{-\gamma }R,  \tag{7}
\end{equation}%
gives

\begin{equation}
\begin{array}{c}
S=\frac{(d-2)}{2}\int_{M^{d}}\sqrt{\tilde{\gamma}}[\Omega ^{-1}\{(\varphi
^{(d-2)}\mathcal{F})((d-3)\frac{\dot{\varphi}^{2}}{\varphi ^{2}}+2\frac{%
\mathcal{\dot{F}}}{\mathcal{F}}\frac{\dot{\varphi}}{\varphi })\} \\ 
\\ 
+\Omega \{k(d-3)\mathcal{F}\varphi ^{(d-4)}\}].%
\end{array}
\tag{8}
\end{equation}%
Here, we used the notation $\mathcal{F}\equiv e^{\frac{f}{2}}$ and $\Omega
\equiv e^{\frac{h}{2}}$. Note that the case $d=2$ is exceptional. Similar
conclusion can be obtained in the case of $d=3$. For our purpose it turns
out convenient to assume that $d-2\neq 0$ and $d-3\neq 0$. Observe that in
(8) $\Omega $ acts as auxiliary field.

Performing variations of the action (8) with respect to $\mathcal{F}$, $%
\Omega $ and $\varphi $ one derives the following equations

\begin{equation}
\frac{2\ddot{\varphi}}{\varphi }+(d-3)\frac{\dot{\varphi}^{2}}{\varphi ^{2}}%
-2\frac{\dot{\Omega}}{\Omega }\frac{\dot{\varphi}}{\varphi }-k(d-3)\Omega
^{2}\varphi ^{-2}=0,  \tag{9}
\end{equation}

\begin{equation}
(d-3)\frac{\dot{\varphi}^{2}}{\varphi ^{2}}+2\frac{\mathcal{\dot{F}}}{%
\mathcal{F}}\frac{\dot{\varphi}}{\varphi }-k(d-3)\Omega ^{2}\varphi ^{-2}=0 
\tag{10}
\end{equation}%
and

\begin{equation}
\begin{array}{c}
\frac{(d-4)(d-3)}{2}\frac{\dot{\varphi}^{2}}{\varphi ^{2}}-\frac{\mathcal{%
\dot{F}}}{\mathcal{F}}\frac{\dot{\Omega}}{\Omega }+\frac{\mathcal{\ddot{F}}}{%
\mathcal{F}}-(d-3)((\frac{\dot{\Omega}}{\Omega }-\frac{\mathcal{\dot{F}}}{%
\mathcal{F}})\frac{\dot{\varphi}}{\varphi }-\frac{\ddot{\varphi}}{\varphi })
\\ 
\\ 
-\frac{k(d-4)(d-3)}{2}\Omega ^{2}\varphi ^{-2}=0,%
\end{array}
\tag{11}
\end{equation}%
respectively. Using (1)-(4) and considering the gravitational field equations

\[
R_{\mu \nu }-\frac{1}{2}\gamma _{\mu \nu }R=0,
\]%
which can be obtained from the action (7), one can verify that the formulae
(9)-(11) are consistent. In particular one can combine (9) and (10) to obtain%
\begin{equation}
\frac{\mathcal{\dot{F}}}{\mathcal{F}}+\frac{\dot{\Omega}}{\Omega }=\frac{%
\ddot{\varphi}}{\dot{\varphi}}.  \tag{12}
\end{equation}%
One recognizes in this formula the typical relation between $\mathcal{F}%
,\Omega $ and $\varphi $ obtained after setting $R_{11}=0$ and $R_{22}=0$ in
(2) and (3) and making the combination $e^{-f+h}R_{11}+R_{22}=0$.

We shall assume that one may associate with (8) the expression (see Refs.
[1]-[5] and references therein)

\begin{equation}
\begin{array}{c}
\mathcal{L}=\frac{1}{2}[\Omega ^{-1}(\varphi ^{(d-2)}\mathcal{F})\{(d-3)%
\frac{\dot{\varphi}^{2}}{\varphi ^{2}}+2\frac{\mathcal{\dot{F}}}{\mathcal{F}}%
\frac{\dot{\varphi}}{\varphi }\} \\ 
\\ 
+\Omega \{k(d-3)\mathcal{F}\varphi ^{(d-4)}\}].%
\end{array}
\tag{13}
\end{equation}%
The object $\mathcal{L}$ is a kind of Lagrangian, but one should keep in
mind that there is not time dependence in (13). So, rigorously speaking,
(13) is not a Lagrangian. For this reason, and since it was derived from the
action (8), let us call $\mathcal{L}$ in (13) `Lagravity', from `Lagrangian
and gravity' relation. We observe that the variable $\Omega $ plays the role
of Lagrange multiplier. This suggests that we seek a possible analogue of
(13) with constraint Hamiltonian formulation. (in order to be consistent we
shall call the kind of Hamiltonian $\mathcal{H}$ associated with $\mathcal{L}
$ `Hagravity'.) For this purpose let us first introduce the redefinition

\begin{equation}
\lambda \equiv \varphi ^{-(d-2)}\mathcal{F}^{-1}\Omega ,  \tag{14}
\end{equation}%
of the Lagrange multiplier $\Omega $. In terms of $\lambda $ the Lagravity,
(13) reads

\begin{equation}
\mathcal{L}=\frac{1}{2}[\lambda ^{-1}((d-3)\frac{\dot{\varphi}^{2}}{\varphi
^{2}}+2\frac{\mathcal{\dot{F}}}{\mathcal{F}}\frac{\dot{\varphi}}{\varphi }%
)+\lambda (k(d-3)\mathcal{F}^{2}\varphi ^{2(d-3)})].  \tag{15}
\end{equation}%
This can be simplified further by introducing the two coordinates $q^{1}$
and $q^{2}$ in the following form:%
\begin{equation}
\varphi \equiv e^{q^{1}}  \tag{16}
\end{equation}%
and

\begin{equation}
\mathcal{F}\equiv e^{q^{2}}.  \tag{17}
\end{equation}%
In fact, in terms $q^{1}$ and $q^{2}$, one sees that (15) can be written as

\begin{equation}
\mathcal{L}=\frac{1}{2}[\lambda ^{-1}((d-3)(\dot{q}^{1})^{2}+2(\dot{q}^{1})(%
\dot{q}^{2}))+\lambda m_{0}^{2}],  \tag{18}
\end{equation}%
where

\begin{equation}
m_{0}^{2}=k(d-3)e^{2(d-3)q^{1}}e^{2q^{2}}).  \tag{19}
\end{equation}

Before proceeding further, it is convenient to verify that we are in the
right route, by first writing the `Euler-Lagrange' equations associated with
(18). For the coordinate $q^{1}$ we have the equation

\begin{equation}
\frac{d}{dr}(\lambda ^{-1}((d-3)\dot{q}^{1}+\dot{q}^{2})-\lambda
(d-3)m_{0}^{2}=0,  \tag{20}
\end{equation}%
and for $q^{2}$ we get

\begin{equation}
\frac{d}{dr}(\lambda ^{-1}\dot{q}^{1})-\lambda m_{0}^{2}=0.  \tag{21}
\end{equation}%
While for $\lambda $ we find%
\begin{equation}
\lambda ^{-2}((d-3)(\dot{q}^{1})^{2}+2\dot{q}^{1}\dot{q}^{2})-m_{0}^{2}=0. 
\tag{22}
\end{equation}%
The idea now is to show that from these equations one can derive (9)-(11).
For this purpose, one first note that from (20) and (21) one finds

\begin{equation}
\frac{d}{dr}(\lambda ^{-1}\dot{q}^{2})=0.  \tag{23}
\end{equation}%
Thus, by writing (21), (22) and (23) in terms $\mathcal{F},\Omega $ and $%
\varphi $ we obtain

\begin{equation}
(d-3)\frac{\dot{\varphi}^{2}}{\varphi ^{2}}+\frac{\ddot{\varphi}}{\varphi }+%
\frac{\mathcal{\dot{F}}}{\mathcal{F}}\frac{\dot{\varphi}}{\varphi }-\frac{%
\dot{\Omega}}{\Omega }\frac{\dot{\varphi}}{\varphi }-k(d-3)\Omega
^{2}\varphi ^{-2}=0,  \tag{24}
\end{equation}

\begin{equation}
(d-3)\frac{\dot{\varphi}^{2}}{\varphi ^{2}}+\frac{2\mathcal{\dot{F}}}{%
\mathcal{F}}\frac{\dot{\varphi}}{\varphi }-k(d-3)\Omega ^{2}\varphi ^{-2}=0 
\tag{25}
\end{equation}%
and

\begin{equation}
(d-2)\frac{\mathcal{\dot{F}}}{\mathcal{F}}\frac{\dot{\varphi}}{\varphi }-%
\frac{\mathcal{\dot{F}}}{\mathcal{F}}\frac{\dot{\Omega}}{\Omega }+\frac{%
\mathcal{\ddot{F}}}{\mathcal{F}}=0,  \tag{26}
\end{equation}%
respectively. We first note that (25) is just (10). Now, multiplying (24) by 
$(d-3)$ and combining the resultant formula with (25) and (26) one sees that
(11) follows. Finally, it is not difficult to obtain (9) from (24) and (25).

For our goal it turns out convenient to solve directly the equations
(21)-(23). Combining (21) and (22) one formally get

\begin{equation}
\lambda \frac{d}{dr}(\lambda ^{-1}\dot{q}^{1})=(d-3)(\dot{q}^{1})^{2}+2\dot{q%
}^{1}\dot{q}^{2}  \tag{27}
\end{equation}%
or

\begin{equation}
-\frac{\dot{\lambda}}{\lambda }\dot{q}^{1}+\ddot{q}^{1}=(d-3)(\dot{q}%
^{1})^{2}+2\dot{q}^{1}\dot{q}^{2}.  \tag{28}
\end{equation}%
This expression suggests that we define the quantity

\begin{equation}
\Omega \equiv e^{(d-2)q^{1}}e^{q^{2}}\lambda ,  \tag{29}
\end{equation}%
which is, of course, consistent with (14). In terms of $\Omega $, (28) gives

\begin{equation}
(d-2)(\dot{q}^{1})^{2}+\dot{q}^{1}\dot{q}^{2}-\frac{\dot{\Omega}}{\Omega }%
\dot{q}^{1}+\ddot{q}^{1}=(d-3)(\dot{q}^{1})^{2}+2\dot{q}^{1}\dot{q}^{2}. 
\tag{30}
\end{equation}%
Simplifying this equation, one sees that (30) is reduced to

\begin{equation}
(\dot{q}^{2}+\frac{\dot{\Omega}}{\Omega })\dot{q}^{1}=\ddot{q}^{1}+(\dot{q}%
^{1})^{2}  \tag{31}
\end{equation}%
or

\begin{equation}
(\dot{q}^{2}+\frac{\dot{\Omega}}{\Omega })=\frac{\ddot{q}^{1}}{\dot{q}^{1}}+%
\dot{q}^{1}.  \tag{32}
\end{equation}%
But since $q^{1}=\ln \varphi $ and $q^{2}=\ln \mathcal{F}$ one discovers
that (32) can be rewritten as (12). The solution of the formula (32) is

\begin{equation}
(q^{2}+\ln \Omega )=\ln \dot{q}^{1}+q^{1}+\ln \alpha ,  \tag{33}
\end{equation}%
where $\alpha $ is a constant. Notice that (33) can also be written as

\begin{equation}
\mathcal{F}\Omega =\alpha \dot{\varphi}.  \tag{34}
\end{equation}%
On the other that from (23) we must have

\begin{equation}
\lambda =\beta \dot{q}^{2},  \tag{35}
\end{equation}%
where $\beta $ is another constant. In terms of $\Omega ,$ $\mathcal{F}$ and 
$\varphi $ the equation (35) becomes

\begin{equation}
\Omega =\beta \varphi ^{(d-2)}\mathcal{\dot{F}}.  \tag{36}
\end{equation}%
Therefore, using (36) the equation (34) leads to

\begin{equation}
\mathcal{F\dot{F}}=\gamma \frac{\dot{\varphi}}{\varphi ^{(d-2)}},  \tag{37}
\end{equation}%
where $\gamma =\frac{\alpha }{\beta }$. The expression (37) implies the
equation%
\begin{equation}
\frac{d}{dr}(\mathcal{F}^{2})=\frac{d}{dr}(-\frac{2\gamma }{(d-3)\varphi
^{(d-3)}}),  \tag{38}
\end{equation}%
whose solution is

\begin{equation}
\mathcal{F}^{2}=a-\frac{b}{\varphi ^{(d-3)}}.  \tag{39}
\end{equation}%
Here, $a$ is a constant and $b=\frac{2\gamma }{(d-3)}$. Moreover, one can
verify that (39) satisfies (25) if $a=k\alpha ^{2}$. In the Newtonian limit
one can set $a=k$ and $b=\frac{G_{d}M}{2(d-3)c^{2}}$, where $G_{d}$ is the $d
$-dimensional Newton gravitational constant, $M$ is the mass associated with
the black hole and $c$ is the light velocity.

Considering (39) one can also determine $\Omega $. In fact, from (34) we see
that

\begin{equation}
\Omega ^{2}=\frac{\alpha ^{2}\dot{\varphi}^{2}}{\mathcal{F}^{2}}=\frac{%
\alpha ^{2}\dot{\varphi}^{2}}{a-\frac{b}{\varphi ^{(d-3)}}}.  \tag{40}
\end{equation}%
Writing $\mathcal{F}=e^{\frac{f}{2}}$ and $\Omega =e^{\frac{h}{2}}$ one
recognizes in (39) and (40) the traditional solutions of a black hole. What
it is interesting about our formalism is that we obtained these solutions by
using the Lagravity (18) rather than the Einstein-Hilbert field equations.

Another advantage of our formalism is that we can now shed some light on the
meaning of the event horizon. Suppose we set $\varphi =r$. From (40) one
sees that $\Omega $ is singular when $a-\frac{b}{r^{(d-3)}}=0$. This surface
defines the so called event horizon. It has been always argued that this is
not true singularity because one can find another set of coordinates (in
particular the Kruskal-Szekeres coordinates) that avoids such a singularity.
From our formalism, however, $\Omega $ (or $\lambda $) plays a role of a
Lagrange multiplier and therefore from (40) one sees that setting $\varphi =r
$ gives $\dot{\varphi}^{2}=1$ and this is equivalent to fix the gauge
associated with $\Omega $. So, from our perspective the event horizon does
not determine a true singularity because, in relation with the parameter $r$%
, such a surface is gauge dependent.

Now, let us compute the analogue of the `canonical momenta' $p_{1}$ and $%
p_{2}$ associated with the coordinates $q^{1}$ and $q^{2}$, respectively.
From (18) one gets

\begin{equation}
p_{1}=\frac{\partial \mathcal{L}}{\partial \dot{q}^{1}}=\lambda ^{-1}((d-3)%
\dot{q}^{1}+\dot{q}^{2})  \tag{41}
\end{equation}%
and

\begin{equation}
p_{2}=\frac{\partial \mathcal{L}}{\partial \dot{q}^{2}}=\lambda ^{-1}\dot{q}%
^{1}.  \tag{42}
\end{equation}

Considering these results one can now show that (18) can be obtained from
the first order Lagravity

\begin{equation}
\mathcal{L}=\dot{q}^{1}p_{1}+\dot{q}^{2}p_{2}-\frac{\lambda }{2}%
(-(d-3)(p_{2})^{2}+2p_{1}p_{2}-m_{0}^{2}).  \tag{43}
\end{equation}

The Lagravity (18) can also be written as

\begin{equation}
\mathcal{L}=\frac{1}{2}[\lambda ^{-1}(\dot{q}^{a}\dot{q}^{b}\xi
_{ab})+\lambda m_{0}^{2}],  \tag{44}
\end{equation}%
where the metric $\xi _{ab}$ is given by

\begin{equation}
\xi _{ab}=\left( 
\begin{array}{cc}
(d-3) & 1 \\ 
1 & 0%
\end{array}%
\right) .  \tag{45}
\end{equation}%
Similarly, we can write (43) in the following form:%
\begin{equation}
\mathcal{L}=\dot{q}^{a}p_{a}-\frac{\lambda }{2}\mathcal{H}.  \tag{46}
\end{equation}%
Here,

\begin{equation}
\mathcal{H}=\xi ^{ab}p_{a}p_{b}-m_{0}^{2}  \tag{47}
\end{equation}%
with%
\begin{equation}
\xi ^{ab}=\left( 
\begin{array}{cc}
0 & 1 \\ 
1 & -(d-3)%
\end{array}%
\right) .  \tag{48}
\end{equation}%
Note that (48) is the inverse matrix of (45). Of course, $\mathcal{H}$ can
be interpreted as `constraint' Hagravity. In fact, it can be shown in
straightforward way that this constraint satisfies the analogue condition of
a first class constraint and therefore it can be interpreted as the gauge
generator of the variable $r$ (see Ref. [9] and references therein).

It is also interesting to write the second order Lagravity

\begin{equation}
\mathcal{L}=\frac{1}{2}m_{0}(\dot{q}^{a}\dot{q}^{b}\xi _{ab})^{1/2}, 
\tag{49}
\end{equation}%
which, by using the corresponding `Euler-Lagrange' equation for $\lambda $,
can be obtained from (44). It is straightforward to see that the associated
`action'

\begin{equation}
\mathcal{S=}\frac{1}{2}\int drm_{0}(\dot{q}^{a}\dot{q}^{b}\xi _{ab})^{1/2}, 
\tag{50}
\end{equation}%
is invariant under reparamitrazation $r^{\prime }=r^{\prime }(r)$. This
explain the reason for existence of the arbitrary function $\lambda $ and
the Hagravity constraint $\mathcal{H}$.

If the cosmological constant $\Lambda $ is included via the usual extended
action $S=\frac{1}{2}\int_{M^{d}}\sqrt{-\gamma }(R-2\Lambda )$ one can show
that the Lagravity (49) also follows but with the `mass'\ $m_{0}^{2}$ now
given by%
\begin{equation}
m_{0}^{2}=k(d-3)e^{2(d-3)q^{1}}e^{2q^{2}}-\frac{2\Lambda }{d-2}%
e^{2(d-2)q^{1}}e^{2q^{2}}.  \tag{51}
\end{equation}

It is worth mentioning that, in four dimensions, that is in $d=4$, from our
formalism arises an intriguing and fascinating result. In such a case the
metric (45) is reduced to

\begin{equation}
\xi _{ab}=\left( 
\begin{array}{cc}
1 & 1 \\ 
1 & 0%
\end{array}%
\right) .  \tag{52}
\end{equation}%
Of course, the matrix (52) can be diagonalized. We first find the
eigenvalues by computing the determinant

\begin{equation}
\left \vert 
\begin{array}{cc}
1-\phi & 1 \\ 
1 & -\phi%
\end{array}%
\right \vert =\phi ^{2}-\phi -1=0.  \tag{53}
\end{equation}%
Surprisingly, one recognizes in (53) the famous formula which determines the
golden ratio. In fact, the two roots of (53) are

\begin{equation}
\phi _{1}=\frac{1+\sqrt{5}}{2}  \tag{54}
\end{equation}%
and

\begin{equation}
\phi _{2}=\frac{1-\sqrt{5}}{2}.  \tag{55}
\end{equation}%
The formula (54) is the so called golden ratio or golden number (see Refs.
[10]-[13] and references therein).

In terms of $\phi _{1}$ and $\phi _{2}$ the Lagravity (49) becomes

\begin{equation}
\mathcal{L}=\frac{1}{2(\sqrt{5})^{1/2}}m_{0}(\omega ^{a}\omega ^{b}\eta
_{ab})^{1/2},  \tag{56}
\end{equation}%
where

\begin{equation}
\begin{array}{c}
\omega ^{1}=\phi _{1}\dot{q}^{1}+\dot{q}^{2}, \\ 
\\ 
\omega ^{2}=\phi _{2}\dot{q}^{1}+\dot{q}^{2}.%
\end{array}
\tag{57}
\end{equation}%
Here, $\eta _{ab}=diag(1,-1)$. Thus, the expression (56) formally
establishes a connection between the golden ratio and black holes. (It is
worth mentioning that in Ref. [14] a different relation between the golden
ratio and black holes has been found. In fact, in such a reference it is
shown that rotating black holes make a phase transition when the ratio of
the square of its mass to the square of its angular momentum is equal to the
golden ratio.)

Summarizing we have developed a variational method that can help to clarify
the meaning of the black holes solution in any dimension. Our procedure is
based in an analogy with the constraint Hamiltonian formalism. As a reward
for our efforts we found an interesting link between two fascinating
concepts: black holes and the golden ratio.

\bigskip \ 

\noindent \textbf{Acknowledgments: }I would like to thank to E. R. Estrada
and L. A. Beltr\'{a}n for helpful comments. This work was partially
supported by PROFAPI-UAS, 2011.

\smallskip \ .

\end{document}